\documentclass[letterpaper, 10 pt,conference]{ieeeconf}
\usepackage{graphicx,epstopdf,multirow,multicol,booktabs,pgfplots}

\IEEEoverridecommandlockouts

\usepackage{amsmath,bm}
\usepackage{amssymb,amsthm}
\usepackage[T1]{fontenc}
\usepackage[colorlinks,citecolor=blue]{hyperref}
\allowdisplaybreaks
\usepackage{setspace}

\usepackage{amssymb}
\usepackage{amsmath}
\usepackage{url}
\usepackage{eucal}
\usepackage{color}

\usepackage{accents}
\newcommand{\ubar}[1]{\underaccent{\bar}{#1}}
\usepackage{hyperref}
\allowdisplaybreaks

\theoremstyle{remark}

\newtheorem{definition}{Definition}

\pgfplotsset{compat=newest}

\title{Dynamic State and Parameter Estimation for Natural Gas Networks using Real Pipeline System Data}

\author{Kaarthik Sundar$^{*}$ \and Anatoly Zlotnik$^{\dagger}$\thanks{$^{*}$Information Systems and Modeling, Los Alamos National
Laboratory, Los Alamos, New Mexico, USA. E-mail: \texttt{kaarthik@lanl.gov}}\thanks{$^{\dagger}$Applied Mathematics and Plasma Physics, Los Alamos National
Laboratory, Los Alamos, New Mexico, USA. E-mail: \texttt{azlotnik@lanl.gov}}\;
}

\begin{document}


\maketitle
\thispagestyle{empty}
\pagestyle{empty}

\begin{abstract}
We present a method for joint state and parameter estimation for natural gas networks where gas pressures and flows through a network of pipes depend on time-varying injections, withdrawals, and compression controls.  The estimation is posed as an optimal control problem constrained by coupled partial differential equations on each pipe that describe space- and time-dependent density and mass flux.  These are discretized and combined with nodal conditions to give dynamic constraints for posing the estimation as nonlinear least squares problems.  We develop a rapid, scalable computational method for performing the estimation in the presence of measurement and process noise. Finally, we evaluate its effectiveness using a data set from a capacity planning model for an actual pipeline system and a month of time-series data from its supervisory control and data acquisition (SCADA) system.
\end{abstract}

\IEEEpeerreviewmaketitle

\section{Introduction} \label{sec:intro}
Gas-fired electricity generators are rapidly replacing coal and nuclear plants, and now produce over 33\% of power used in the United States. They provide base load, and because of their responsiveness are also used to balance out fluctuations in production by uncontrollable renewables such as wind and solar \cite{Rinaldi2001,Li2008}.  This creates large and rapid changes in natural gas consumption, which impacts gas pressures and flows in the gas networks that deliver fuel to the power plants. These new conditions render the steady-state gas flow models, widely used in methods for estimating capacity of and optimizing natural gas systems \cite{Wong1968,Rothfarb1970}, unusable.

Accurate yet tractable models for dynamic gas flows are critical for state and parameter estimation problems for large gas pipeline networks. Monitoring, leak detection, and predictive simulation all require precise information about the instantaneous network state \cite{Pal1991}, and methods typically focus on detection on a single pipe \cite{aamo2016leak}. In practice, it is too costly to place pressure and flow meters everywhere throughout a pipeline system. There is growing interest in rapid and scalable state and parameter estimation techniques that use transient measurements of pressure and gas-withdrawals obtained at network locations where they are available \cite{Reddy2011}.

Gas flows in a pipeline are represented by a system of coupled partial differential equations (PDEs), namely, the Euler equations in one dimension \cite{Osiadacz1984}. Many methods have been proposed to simulate such flows \cite{brouwer2011}, and research on gas network simulation is still active \cite{wang2018}, yet modeling of gas network flows for estimation problems remains challenging.  Linearization of the PDEs around the steady-state solution was used to obtain transfer function and state space models for gas network dynamics \cite{Kralik1984, Reddy2006, Reddy2011, Alamian2012}, which were used for state estimation with traditional methods for linear systems.  As noted above, the emerging influence of gas-fired power plants causes transient phenomena that force pipeline flows to deviate substantially from steady-state. Engineering limits that require bounds on state and control variables add complex algebraic constraints.  This motivates gas network estimation techniques that use truly transient models, respect constraints on system states and actuation, and which scale to large systems with arbitrary network structures.

In this study we formulate and solve joint state and parameter estimation problems for transient flows in gas networks.  In contrast to approaches that use linearization  \cite{Kralik1984,Reddy2006,Alamian2012}, we approximate the PDE constraints using a  nonlinear control system model derived by reduction of the gas network dynamics \cite{Grundel2013,Zlotnik2015,Zlotnik2015a}.  The resulting implicit nonlinear differential algebraic equation (DAE) system is discretized in time and derivative terms are approximated using finite differences to yield a nonlinear algebraic equation system that approximates gas network dynamics with proven fidelity \cite{Zlotnik2015a,gyrya2019}.  This scheme enables gas network estimation to be formulated as a least squares problem in the form of a nonlinear program (NLPs). 
We evaluate the effectiveness of the methodology, which was derived in recent work by the authors \cite{sundar2018}, by applying it to a test case created using a capacity planning network model for an actual pipeline system and a month of time-series measurements from its supervisory control and data acquisition (SCADA) system, which is unprecedented to our knowledge.

The rest of the article is organized as follows. We review the PDE system for gas pipeline flow and the model for a gas network with controllable compressors and variable injections. We then present a DAE model for the gas network dynamics that is used for computational estimation.  We then state the joint state and parameter estimation problem. This is followed by a description of the real data test case and results of computations done with this data, and then a conclusion.

\section{Modeling} \label{sec:model}
\subsection{Gas pipeline dynamics} \label{subsec:dynamics}
The flow of compressible gas within a horizontal pipe with slow transients that do not excite waves or shocks is adequately described using the one-dimensional Euler equations \cite{Thorley1987}.  We use the simplified system
\begin{flalign}
& \partial_t \rho + \partial_x \varphi = 0 \quad \text{and} \quad a^2 \partial_x \rho = -\frac{\lambda}{2D} \frac{\varphi |\varphi|}{\rho}. & \label{eq:pde_1}
\end{flalign}
The first and second equations above represent conservation of mass and momentum, respectively. The variables $\rho$ and $\varphi$ represent instantaneous gas density and mass flux, respectively, and are defined on the domain $[0,\ell] \times [0,T]$ where $\ell$ represents the length of a pipe. The term on the right hand side of the second equation aggregates friction effects, where the parameters are the Darcy-Wiesbach friction factor $\lambda$ and pipe diameter $D$.  We assume that gas pressure $p$ and density $\rho$ satisfy the ideal equation of state $p = a^2 \rho$ with $a^2 = ZR\boldsymbol{T}$, where $a$, $Z$, $R$, and $\boldsymbol{T}$, are the speed of sound, gas compressibility factor, ideal gas constant, and constant temperature, respectively.  The Equations \eqref{eq:pde_1} are valid in the regime when changes in the boundary conditions are sufficiently slow to not excite propagation of sound waves \cite{Chertkov2015}. Multiple studies have supported the use of this simplification \cite{Osiadacz1984,herty2010}, and a detailed justification for its use in gas network estimation problems is available \cite{sundar2018}.

Eq. \eqref{eq:pde_1} has a unique solution when the initial conditions and boundary conditions consisting of one of $\rho(0,t) = \ubar{\rho}(t)$ or $\varphi(0,t) = \ubar{\varphi}(t)$ and one of $\rho(\ell,t) = \bar{\rho}(t)$ and $\varphi(\ell,t) = \bar{\varphi}(t)$ are specified for the pipe. For both convenience and to improve numerical conditioning of Eq. \eqref{eq:pde_1}, we apply the dimensional transformations
\begin{flalign} \label{eq:nondim}
&\hat{t}=\frac{t}{\ell_0/a}, \quad \hat{x}=\frac{x}{\ell_0}, \quad \hat{\rho}=\frac{\rho}{\rho_0}, \quad \hat{\varphi}=\frac{\varphi}{a\rho_0},&
\end{flalign}
where $\ell_0$ and $\rho_0$ are nominal length and density, to yield the non-dimensional gas dynamics on a pipeline,
\begin{flalign}
& \partial_t \rho + \partial_x \varphi = 0 \quad \text{and} \quad \partial_x \rho = -\frac{\lambda \ell_0}{2D} \frac{\varphi |\varphi|}{\rho}. & \label{eq:pde_nondim}
\end{flalign}
The hat symbols in the above non-dimensional equations have been omitted for readability. 

The friction of turbulent flow in each pipe causes pressure in the pipeline to gradually decrease along the direction of flow. Gas compressors are used to boost line pressure to meet the minimum pressure requirement for delivery to customers.  We model compressor stations as controllers that manipulate the state of the gas transmission system by changing the density ratio between outlet and inlet. The compressor station is small relative to the length of a pipeline, so we represent compressor action as a multiplicative increase in the density at a point $x = c$ with conservation of flow \textit{i.e.}, $\rho(c^+, t) = \alpha(t) \cdot \rho(c^-,t)$ and $\varphi(c^+,t) = \varphi(c^-,t)$ where, $\alpha(t)$ denotes the time-dependent compression ratio between suction (intake) and discharge (outlet) pressure.

\subsection{Dynamics of gas network flow} \label{subsec:network_model}
A gas network consists of pipes (edges) interconnected at junctions (nodes) where the gas flow can be compressed, withdrawn from, or injected into the system.  We model the gas pipeline network as a connected directed graph $\mathcal G = (\mathcal V, \mathcal E)$ where $\mathcal V$ and $\mathcal E$ represent the set of junctions and the set of pipelines connecting any pair of junctions, respectively. We use $(i,j) \in \mathcal E$ to denote the pipeline that connects the junctions $i, j\in \mathcal V$. Let $\rho_{ij}$ and $\varphi_{ij}$ denote the instantaneous density and mass flux, respectively, within the edge $(i,j) \in \mathcal E$ defined on the domain $[0, L_{ij}] \times [0,T]$.  Each pipe $(i,j)$ is characterized by its length $L_{ij}$, diameter $D_{ij}$, and friction factor $\lambda_{ij}$. The cross-sectional area of each pipe is denoted by $X_{ij}$. Between any two junctions that are connected via a pipe, the mass flux and density evolve according to Eq. \eqref{eq:pde_nondim}.  Hence for each edge $(i,j)\in \mathcal E$, the evolution of $\rho_{ij}$ and $\varphi_{ij}$ is given by Eq. \eqref{eq:pde_nondim}, \emph{i.e.},
\begin{flalign}
& \partial_t \rho_{ij} + \partial_x \varphi_{ij} = 0 \,\, \text{and} \,\, \partial_x \rho_{ij} = -\frac{\lambda_{ij} \ell_0}{2D_{ij}} \frac{\varphi_{ij} |\varphi_{ij}|}{\rho_{ij}} & \label{eq:pde_nondim_edge_both}
\end{flalign}

\begin{figure}[t!]
\centering
\includegraphics[scale=1]{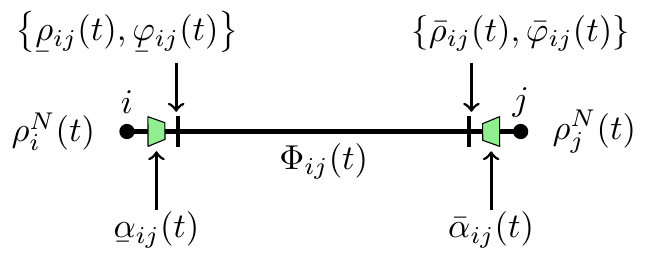}
\caption{The figure shows the densities and flows at the boundaries of each edge and the compression that can be applied at both the nodes $i$ and $j$.}
\label{fig:schematic}
\end{figure}

Here $\varphi_{ij}$ is directional, where the sign indicates flow direction. We use a directed graph, which leads to $\varphi_{ij}(x_{ij},t)=-\varphi_{ji}(L_{ij}-x_{ij},t)$. In addition, every junction $i \in \mathcal V$ is associated with a time-dependent nodal density $\rho^N_i(t): [0,T] \to \mathbb R_+$. The set of controllers is denoted by $\mathcal C\subset \mathcal E\times\{+,-\}$, where $(i,j)\equiv(i,j,+)\in\mathcal C$ denotes a controller located at node $i\in\mathcal V$ that augments the density of gas flowing into edge $(i,j)\in\mathcal E$ in the direction $i \to j$, while $(j,i)\equiv(i,j,-)\in\mathcal C$ denotes a controller located at node $j\in\mathcal V$ that augments density into edge $(i,j)\in\mathcal E$ in the direction $j\to i$. Compression is then modeled as a multiplicative ratio $\ubar{\alpha}_{ij}:[0,T]\to\mathbb R_+$ for $(i,j, +)\in\mathcal C$ and $\bar{\alpha}_{ij}:[0,T]\to\mathbb R_+$ for $(i,j, -)\in\mathcal C$.  Now let $\mathcal V_s \subset \mathcal V$ denote the set of supply junctions where gas enters the network. Let $s_j(t)$ be the time-varying supply density at the junction $j \in \mathcal V_s$. Mass flux withdrawals at the remaining junctions $j \in \mathcal V_d = \mathcal V \setminus \mathcal V_s$ are denoted by $d_j(t)$. For ease of exposition, we shall refer to the $\mathcal V_s$ and $\mathcal V_d$ as the set of ``slack'' and ``non-slack'' nodes, respectively. Nodal balance equations characterize the boundary conditions for the dynamics in Eq. \eqref{eq:pde_nondim_edge_both}. We define densities and flows at edge domain boundaries by
\begin{subequations}
\begin{flalign}
& \ubar{\rho}_{ij}(t) \triangleq \rho_{ij}(t, 0), \quad \bar{\rho}_{ij}(t) \triangleq \rho_{ij}(t, L_{ij}), & \label{eq:rhobar} \\
& \ubar{\varphi}_{ij}(t) \triangleq \varphi_{ij}(t, 0), \quad \bar{\varphi}_{ij}(t) \triangleq \varphi_{ij}(t, L_{ij}), & \label{eq:phibar}\\
& \text{and the nominal average edge flow as} \nonumber \\
& \Phi_{ij}(t) \triangleq \frac 12 (\ubar{\varphi}_{ij}(t) + \bar{\varphi}_{ij}(t))). &\label{eq:avg_phi}
\end{flalign}
\label{eq:bar_defn}
\end{subequations}
For ease of understanding, the above definitions are illustrated using the schematic in Fig. \ref{fig:schematic}, on a pipe joining two nodes $i$ and $j$.
Nodal balance laws are specified in terms of time-dependent compressor ratios $\ubar{\alpha}_{ij}(t)$ and $\bar{\alpha}_{ij}(t)$, gas withdrawals $d_j(t)$, and supply densities $s_j(t)$ as
\begin{subequations}
\begin{align}
\!\!\!\!\!\! \ubar{\rho}_{ij}(t) &= \ubar{\alpha}_{ij}(t)\rho_i^N(t), \, \forall \, (i,j) \in \mathcal E,  \label{eq:nodal_density_balance_1}\\
\!\!\!\!\!\! \bar{\rho}_{ij}(t) &= \bar{\alpha}_{ij}(t)\rho_i^N(t), \, \forall \, (i,j) \in \mathcal E,  \label{eq:nodal_density_balance_2}\\
\!\!\!\!\!\! d_j(t) &=\sum_{i\in\mathcal V_d}X_{ij} \bar{\varphi}_{ij}(t)- \sum_{k\in\mathcal V_d}X_{jk}\ubar{\varphi}_{jk}(t), \,\forall\, j\in\mathcal V_d,  \label{eq:flow_balance} \\
\!\!\!\!\!\! \ubar{\rho}_{ij}(t) &= s_i(t), \,\forall\, i\in\mathcal V_s.  \label{eq:slack_pressure}
\end{align}
\label{eq:nodal_balance}
\end{subequations}
We suppose that measurements of the time-varying compressor ratio functions $\{\ubar{\alpha}_{ij}, \bar{\alpha}_{ij}\}_{(i,j)\in\mathcal C}$ and the transient withdrawals $\{d_j\}_{j\in\mathcal V_d}$ are available a priori.
Time-varying pressures at slack nodes are also assumed known. In the rest of the article, we shall not show the explicit dependence of the density and flow variables on the independent variable $t$.

We suppose that densities of gas in a pipe are upper and lower bounded according to the constraints
\begin{flalign}
& \rho_{ij}^{\min}\leqslant \rho_{ij}(t,x_{ij}) \leqslant \rho_{ij}^{\max}, \,\forall\, (i,j)\in\mathcal E. &\label{eq:box_density}
\end{flalign}
For the time horizon $T$ (significant transients of interest typically occur over $T=24$ hours), we require that the state variables $\varphi_{ij}$ and $\rho_{ij}$ are time-periodic in order for the dynamic constraints to be well-posed, and time-periodicity also has to be imposed on the control and parameter functions $\{\ubar{\alpha}_{ij}, \bar{\alpha}_{ij}\}_{(i,j)\in\mathcal C}$, $\{d_j\}_{j\in\mathcal V_d}$, and $\{s_j\}_{j\in\mathcal V_s}$ as given by Eqs. \eqref{eq:termcon1c}--\eqref{eq:termcon1e} (see \cite{Zlotnik2015}).
This yields terminal conditions
\begin{subequations}
\begin{flalign}
&\rho_{ij}(0,x_{ij})=\rho_{ij}(T,x_{ij}),  \,\forall\, (i,j)\in\mathcal E, &\label{eq:termcon1a}\\
&\phi_{ij}(0,x_{ij})=\phi_{ij}(T,x_{ij}),  \,\forall\, (i,j)\in\mathcal E, &\label{eq:termcon1b}\\
&\ubar{\alpha}_{ij}(0)=\ubar{\alpha}_{ij}(T), \, \bar{\alpha}_{ij}(0)=\bar{\alpha}_{ij}(T),  \,\forall\, (i,j)\in\mathcal C, &\label{eq:termcon1c} \\
&d_j(0)= d_j(T), \,\forall\, j \in \mathcal V_d, &\label{eq:termcon1d} \\
&s_j(0)=s_j(T), \,\forall\, j \in \mathcal V_s. &\label{eq:termcon1e}
\end{flalign}
\label{eq:termcon}
\end{subequations}
The primary justification for using time-periodic boundary conditions is the need for well-posedness, both conceptually and computationally. Conceptually, without some specification of the initial and terminal conditions, these states could be produced by the solver in unpredictable ways.  
Computationally, time-periodicity can be implemented by using a single vector to store the initial and terminal points in the discretization.   This ``circular'' time-discretization reduces the problem size and eliminates the need for additional constraints on the initial and terminal states.  For assimilating data that is not time-periodic, the estimation problem can be applied to an extended time-horizon over which data are interpolated to produce periodic inputs, and the solution can be taken as the restriction to the time-horizon of interest.


\subsection{Control system model} \label{subsec:cs_model}
In this section, we develop a reduced order model that represents the dynamics of gas flow through a network using a synthesis of Eqs. \eqref{eq:pde_nondim_edge_both}, \eqref{eq:nodal_balance}, \eqref{eq:box_density}, and \eqref{eq:termcon}. A lumped element approximation is made to characterize the dynamics for each edge in Eq. \eqref{eq:pde_nondim_edge_both}, together with Eq. \eqref{eq:nodal_balance} and Eq. \eqref{eq:box_density}, that uses nodal density $\rho_i^N$ for every $i \in \mathcal V$ as the state of the system. This reduction extends the previous modeling work  \cite{Grundel2013,Zlotnik2015,Zlotnik2016}.
To that end, we introduce definitions used in \cite{Zlotnik2016}.
\begin{definition} \textit{Spatial Graph Refinement}: A refinement $\hat{\mathcal G} = (\hat{\mathcal V}, \hat{\mathcal E})$ of a directed graph $\mathcal G = (\mathcal V, \mathcal E)$ with a length $L_{ij}$ associated with each edge $(i,j) \in \mathcal E$ is constructed by adding extra nodes to subdivide the edges of $\mathcal E$ such that, the length of a new edge $(i,j) \in \hat{\mathcal E}$, $\hat L_{ij}$ satisfies
\begin{flalign}
& \frac{\Delta L_{\mu(ij)}}{\Delta + L_{\mu(ij)}} < \hat L_{ij} < \Delta, & \label{eq:refinement}
\end{flalign}
where, $\mu:\hat{\mathcal E} \to \mathcal E$ is a surjection from the refined edges to the parent edges and $\Delta$ denotes the maximum edge length in the refined graph produced by spatial discretization.
\end{definition}
When refining the graph $\mathcal G$, we assume that $L = \Delta$ is small, so the relative difference of the density and mass flux at the start and end of this refined edge is small \emph{i.e.},
\begin{flalign}
& \frac{\bar{\rho}_{ij} - \ubar{\rho}_{ij}}{\bar{\rho}_{ij} + \ubar{\rho}_{ij}} \ll 1, \quad\text{and}\quad \frac{\bar{\varphi}_{ij} - \ubar{\varphi}_{ij}}{\bar{\varphi}_{ij} + \ubar{\varphi}_{ij}} \ll 1 \, \forall \, (i,j) \in \hat{\mathcal E}. & \label{eq:lumping_condition}
\end{flalign}
With this assumption, the relative density difference between the neighboring nodes is minor at all times. The dynamics for each pipeline segment $(i,j) \in \hat{\mathcal E}$ in the refined graph $\hat{\mathcal G} = (\hat{\mathcal V}, \hat{\mathcal E})$ is again given by \eqref{eq:pde_nondim_edge_both}, and lumping yields
\begin{subequations}
\begin{flalign}
& \int_0^L(\partial_t\rho_{ij}+\partial_x\varphi_{ij}) \,dx=0, & \label{eq:mass_int} \\
& \int_0^L(\partial_x \rho_{ij})\,dx = -\frac{\lambda_{ij}\ell_0}{2D_{ij}}\int_0^L \frac{\varphi_{ij} |\varphi_{ij}|}{\rho_{ij}}\,dx. & \label{eq:momentum_int}
\end{flalign}
\label{eq:pde_int}
\end{subequations}
The above integrals of $\partial_t$, $\partial_x$, and nonlinear terms are evaluated using the trapezoid rule, the fundamental theorem of calculus, and averaging variables, respectively, yielding
\begin{subequations}
\begin{flalign}
& \frac{L}{2}(\dot{\ubar{\rho}}_{ij}+\dot{\bar{\rho}}_{ij}) = \ubar{\varphi}_{ij}-\bar{\varphi}_{ij}, & \label{eq:mass_disc}\\
& \ubar{\rho}_{ij}-\bar{\rho}_{ij} = -\frac{\lambda_{ij}\ell_0 L}{4D_{ij}} \frac{(\ubar{\varphi}_{ij}+\bar{\varphi}_{ij}) |\ubar{\varphi}_{ij}+\bar{\varphi}_{ij}|}{\ubar{\rho}_{ij}+\bar{\rho}_{ij}}. & \label{eq:momentum_disc}
\end{flalign}
\label{eq:ode_disc}
\end{subequations}
The equations \eqref{eq:ode_disc} and nodal balance laws \eqref{eq:nodal_balance} then reduce to a differential-algebraic equation (DAE) system:
\begin{subequations}
\begin{flalign}
& \frac{L}{2}(\dot{\ubar{\rho}}_{ij}+\dot{\bar{\rho}}_{ij}) = \ubar{\varphi}_{ij}-\bar{\varphi}_{ij},\, \forall \, (i,j) \in \hat{\mathcal E}& \label{eq:dae0c}\\
& \ubar{\rho}_{ij}-\bar{\rho}_{ij} = -\frac{\lambda_{ij}\ell_0 L}{D_{ij}} \frac{\Phi_{ij}|\Phi_{ij}|}{(\ubar{\rho}_{ij}+\bar{\rho}_{ij})},  \, \forall \, (i,j) \in \hat{\mathcal E} & \label{eq:dae0d} \\
& \ubar{\rho}_{ij} = \ubar{\alpha}_{ij}\rho_i^N, \,\bar{\rho}_{ij} = \bar{\alpha}_{ij}\rho_i^N, \, \forall \, (i,j) \in \hat{\mathcal E}, &\label{eq:dae0a}\\
& d_j = \sum_{i\in\hat{\mathcal V}_d}X_{ij} \bar{\varphi}_{ij} - \sum_{k\in\hat{\mathcal V}_d}X_{jk}\ubar{\varphi}_{jk}, \, \forall \, j\in\hat{\mathcal V}_d,  \label{eq:dae0b} & \\
& \ubar{\rho}_{ij} = s_i, \,\forall \, i\in\hat{\mathcal V}_s. & \label{eq:dae0e}
\end{flalign}
\label{eq:dae}
\end{subequations}
Eq. \eqref{eq:dae0a} represents continuity of density at junctions with jumps in the case of compression or regulation, Eq. \eqref{eq:dae0b} represents flow balance at junctions, and Eqs. \eqref{eq:dae0c}-\eqref{eq:dae0d} represent flow dynamics on each segment.

The DAE system in \eqref{eq:dae} can be written as a system of nonlinear DAEs in matrix-vector form using graph theoretic notation. We enumerate the set of nodes in the set $\hat{\mathcal V}$ according to a fixed ordering, where the non-slack nodes $\hat{\mathcal V}_d$ are ordered after the slack nodes, $\hat{\mathcal V}_s$. Now each node in $\hat{\mathcal V}$ is assigned an index $[\hat{\mathcal V}] := \{1,\dots, |\hat{\mathcal V}|\}$ according to the chosen ordering. Each edge is also assigned an index in $[\hat{\mathcal E}] := \{1, \dots, |\hat{\mathcal E}|\}$ and we define the map $\pi_e:\hat{\mathcal E} \to [\hat{\mathcal E}]$, which maps each edge to this ordering. In the rest of the article, boldface notation will be used to represent vectors.

Let $\bm \rho^N = (\rho_1^N, \rho_2^N, \dots, \rho_{|\hat{\mathcal V}|}^N)^{\intercal}$ denote the nodal density state vector.
Equation \eqref{eq:dae0a} will be used to state \eqref{eq:dae0c}-\eqref{eq:dae0d} in terms of nodal densities $\bm \rho^N$.  We then define state vectors $\ubar{\bm \varphi}=(\ubar{\varphi}_1,\ldots,\ubar{\varphi}_{|\hat{\mathcal E}|})^\intercal$ and $\bar{\bm \varphi}=(\bar{\varphi}_1,\ldots,\bar{\varphi}_{|\hat{\mathcal E}|})^\intercal$, where $\ubar{\varphi}_k$ and $\bar{\varphi}_k$ are indexed by $k=\pi_e(ij)$. Furthermore, we let $\bm \Phi = \tfrac 12 (\ubar{\bm \varphi} + \bar{\bm \varphi})$ denote the vector of average flow in each pipeline segment.
We now define the incidence matrix of the full refined graph $\hat{\mathcal G}$, acting
$A:\mathbb R^{|\hat{\mathcal E}|}\to\mathbb R^{|\hat{\mathcal V}|}$, by
\begin{flalign} \label{eq:incidence0}
&A_{ik} = \left\{ \begin{array}{ll}  1 & \text{edge $k=\pi_e(ij)$ enters node $i$,} \\ -1 & \text{edge $k=\pi_e(ij)$ leaves node $i$,} \\ 0 & \text{else} \end{array}\right. &
\end{flalign}
and also the time-dependent weighted incidence matrix $B:\mathbb R^{|\hat{\mathcal E}|}\to\mathbb R^{|\hat{\mathcal V}|}$ given by
\begin{flalign} \label{eq:incidence0_w}
&B_{ik} = \left\{ \begin{array}{ll}  \bar{\alpha}_{ij} & \text{edge $k=\pi_e(ij)$ enters node $i$,} \\ -\ubar{\alpha}_{ij} & \text{edge $k=\pi_e(ij)$ leaves node $i$,} \\ 0 & \text{else}, \end{array}\right. &
\end{flalign}
where $\operatorname{sign}(B)=A$.  Here the compressor controls are embedded within the matrix $B$. A vector of withdrawal fluxes is defined by $\bm d=(d_1,\ldots,d_M)^T$ with $M=|\hat{\mathcal V}_d|$, where $d_k$ is negative if an injection. We also define the slack node densities as $\bm s=(s_1,\ldots,s_b)^{\intercal}=\{\rho^N_j\}_{j\in\hat{\mathcal V}_s}$, where $b=|\hat{\mathcal V}_s|$, and non-slack (demand) node densities as $\bm \rho=(\rho_1,\ldots,\rho_M)^T=\{\rho^N_j\}_{j\in\hat{\mathcal V}_d}$, so that $b+M=|\hat{\mathcal V}|$. Note that $\bm s$, $\bm \rho$ and $\bm \rho^N$ are related by $\bm \rho^N = (\bm s, \bm \rho)^\intercal$, because of the choice of node ordering $\hat{\mathcal V}$. We let $A_s,B_s\in\mathbb R^{b\times |\hat{\mathcal E}|}$ denote the sub-matrices of rows of $A$ and $B$ corresponding to $\hat{\mathcal V}_s$, and let $A_d,B_d\in\mathbb R^{M\times |\hat{\mathcal E}|}$ correspond similarly to $\hat{\mathcal V}_d$.
We then define the diagonal matrices $\Lambda,K, \boldsymbol{X}\in\mathbb R^{|\hat{\mathcal E}|\times |\hat{\mathcal E}|}$ by $\Lambda_{kk}=L_k$,  $K_{kk}=\ell_0\lambda_k/D_k$, and $\boldsymbol{X}_{kk} = X_k$ where $L_k$, $\lambda_k$, $D_k$, and $A_k$ are the non-dimensional length, friction factor, diameter, and cross-sectional area of edge $k=\pi_e(ij)$. Using this notation, \eqref{eq:dae} can be rewritten as a DAE system:
\begin{subequations}
\begin{flalign}
& |A_d| \boldsymbol{X} \Lambda |B_d^\intercal|\dot{\bm \rho} = 4(A_d \boldsymbol{X} \bm \Phi - \bm d) - |A_d| X \Lambda |B_s^\intercal| \dot{\bm s},  & \label{eq:dae1a} \\
& \Lambda K \bm \Phi \odot \bm \Phi =  -B^\intercal \bm \rho^N \odot |B^\intercal| \bm \rho^N, & \label{eq:dae1b}
\end{flalign}
\label{eq:dae_final}
\end{subequations}
\noindent where $\odot$ represents the Hadamard product.   Here, gas withdrawals are $\bm d \in \mathbb R^M$, input densities are $\bm s \in \mathbb R_+^b$ and the compression ratios $\ubar{\alpha}_{ij}, \bar{\alpha}_{ij} \in \mathcal C$ are time-varying and $\bm \rho \in \mathbb R_+^M$ and $\bm \Phi \in \mathbb R^{|\hat{\mathcal E}|}$ denote the states of the system.

To see this, we rewrite Eq. \eqref{eq:dae0b} in matrix form as $\bm d = \bar{A}_d \boldsymbol{X} \bar{\bm \varphi} + \ubar{A}_d \boldsymbol{X} \ubar{\bm \varphi}$
where $\bar{A}_d$ and $\ubar{A}_d$ are the positive and negative parts of the matrix $A_d$, respectively. We now define $\bm \Phi_{-} = \frac 12 (\bar{\bm \varphi} - \ubar{\bm \varphi})$.  The Eq. \eqref{eq:dae0b} can then be rewritten as in the transformed variables $\bm \Phi$ and $\bm \Phi_{-}$ as
\begin{flalign}
& \bm d = A_d \boldsymbol{X} \bm \Phi + |A_d| \boldsymbol{X} \bm \Phi_{-}. & \label{eq:d_veca}
\end{flalign}
On the other hand, Eqs. \eqref{eq:dae0a}, \eqref{eq:dae0c}, and \eqref{eq:dae0e} together with the definition $\bm \Phi_{-}$ can be equivalently represented using the matrix equation
\begin{flalign}
& |B_s^\intercal| \dot{\bm s} + |B_d^\intercal| \dot{\bm \rho} = -4 \Lambda^{-1} \bm \Phi_{-}.  & \label{eq:rho_dynamics_vec}
\end{flalign}
Substituting Eq. \eqref{eq:d_veca} into  \eqref{eq:rho_dynamics_vec} and eliminating $\bm \Phi_{-}$ yields \eqref{eq:dae1a}. Eq. \eqref{eq:dae0d} can be rewritten as \begin{flalign}
& \ubar{\rho}_{ij}^2 - \bar{\rho}_{ij}^2 = -\frac{\lambda \ell_0 L}{D_{ij}} \Phi_{ij}|\Phi_{ij}|, \, \forall (i,j) \in \hat{\mathcal E}. & \label{eq:phi_vec}
\end{flalign}
Using Eq. \eqref{eq:dae0c} and the definitions of $B$, $\Lambda$, and $K$, the above equation can be  written in matrix form as \eqref{eq:dae1b}.

\subsection{Uncertainty modeling} \label{subsec:noise}
We model uncertainty in the dynamics in Eq. \eqref{eq:dae_final} by incorporating an additive noise process $\bm \eta$ to Eq. \eqref{eq:dae1a}:
\begin{align}
|A_d| \boldsymbol{X} \Lambda \left( |B_d^\intercal|\dot{\tilde{\bm \rho}} + |B_s^\intercal| \dot{\bm s} \right) \! + \! 4(-A_d \boldsymbol{X} \bm \Phi \!+\! \tilde{\bm d}) \! + \! \bm \eta \! = \! \bm 0  & \label{eq:ode_nodal_noise}
\end{align}
In the above equation, $\bm \eta$ has a dependence on time that has not been made explicit for sake of readability.  We use the noise process $\bm \eta$ to simultaneously account for errors caused by (i) simplification of physical modeling, (ii) uncertainty in model parameters, and (iii) process and measurement noise.  Specifically, the value $\tilde{\bm \rho}$ represents the solution to the stochastic DAE in \eqref{eq:ode_nodal_noise} given stochastic withdrawals $\tilde{\bm d}$.  For the estimation problems, we assume that  $\tilde{\bm d}$ and $\tilde{\bm \rho}$ are available, and we interpret them as noisy measurements of $\bm d$ and $\bm \rho$, which in turn satisfy the deterministic (noiseless) model \eqref{eq:dae_final}.   After time-discretization of the system \eqref{eq:ode_nodal_noise} the process $\bm \eta$ can be interpreted to incorporate error caused by coarse sampling in time.  
Throughout the rest of the article we refer without loss of generality to $\bm \eta$ as the measurement noise.  We do not make any other assumption on $\bm \eta$.  In Section \ref{sec:state_param}, we formulate least squares problems for weighted $L_2$ minimization of the measurement and process errors over a time horizon $T$, i.e., $\int_0^T(\bm d-\tilde{\bm d})^{\intercal} W_1(\bm d-\tilde{\bm d}) \mathrm{d}t$ and $\int_0^T(\bm \rho - \tilde{\bm \rho})^\intercal W_2(\bm \rho - \tilde{\bm \rho}) \mathrm{d}t$, where $W_1$ and $W_2$ are the respective weighting matrices, subject to the deterministic dynamic constraints \eqref{eq:dae_final}.  
Our purpose is to develop an applied technique for state and parameter estimation for pipeline system models of the form \eqref{eq:dae_final}, so we do not attempt to analyze the characteristics of $\bm \eta$ here.

\section{Estimation Formulation} \label{sec:state_param}
The formulation for the joint state and parameter estimation problem is given as a nonlinear least squares problem where uncertain time-varying gas withdrawals, nodal pressures, and compressor ratios are known a priori.  All measurements are assumed to be obtained only from the physical nodes, $\mathcal V$, in the graph $\mathcal G$.

In practice, the gas withdrawal profiles are measured using flow meters and are noisy or uncertain. Additionally, the pressure gauges at the junctions provide noisy nodal pressure measurements that can be converted to noisy density measurements. 
Let $\tilde{d}_j(t)$, $j \in \mathcal V_D$ denote these measured withdrawal profiles and $\tilde{\rho}_j(t)$, $j \in \mathcal V_D$ denote the density (pressure) measurements at the non-slack nodes; let $\tilde{\bm d}$ and $\tilde{\bm \rho}$ represent the corresponding vector of measurements. 
We assume that the slack nodal pressure vector is known. For the joint state  and parameter estimation problem, the variables in the nonlinear least squares formulation are $\bm \rho$, $\bm \Phi$, $\bm d$, and $K$ in \eqref{eq:dae1b}, which contains the friction factor of each pipe $(i,j)$ (denoted by $\lambda_{ij}$).
We formulate a weighted nonlinear least squares problem using a running cost objective function:

\begin{flalign}
\mathcal{L}(\bm d,\tilde{\bm d}, \bm \rho,\tilde{ \bm \rho}) \equiv & \int_0^T(\bm d - \tilde{\bm d})^\intercal W_1 (\bm d - \tilde{\bm d}) \nonumber \\  & \qquad +  (\bm \rho - \tilde{ \bm \rho})^\intercal W_2 (\bm \rho - \tilde{ \bm \rho}) \mathrm{d}t & \label{eq:state_2_obj}
\end{flalign}

Then, the estimation problem is formulated as
\begin{subequations}
\begin{flalign}
    & \min_{\bm \rho,\,\bm\Phi,\,\bm d,\,K} \, \mathcal{L}(\bm d,\tilde{\bm d}, \bm \rho,\tilde{ \bm \rho}) \notag \\ 
    & \text{subject to: } \text{Eqs. \eqref{eq:dae_final}, } & \notag \\
    & {\bm \rho}^{\min} \leqslant {\bm \rho} \leqslant {\bm \rho}^{\max}, & 
     \label{eq:state_2_bounds_sp} \\
    & {\bm \rho}(0) = {\bm \rho}(T), \, {\bm \Phi}(0) = {\bm \Phi}(T), \text{ and } {\bm d}(0) = {\bm d}(T).  &\label{eq:state_2_periodicity_sp}
\end{flalign}
\label{eq:state_2_sp}
\end{subequations}
The optimized variables are densities $\bm \rho$, per-area mass flows $\bm \Phi$, and estimated withdrawals $\bm d$, as well as friction factor parameters $\lambda_{ij}$ for each edge $\mathcal{E}$ in the original physical graph $\mathcal{G}$.  The constraints in Eq. \eqref{eq:state_2_bounds_sp} impose bounds on the nodal density variables. The above formulation computes time-periodic estimates of the state variables and the withdrawals. We justify this least-squares method by the principle of convergence in the limit as the magnitude of noise decreases to zero.  That is, the state estimates approach the actual state as measurement noise $\tilde{d}$ is decreased.  We use a finite difference approximation for the derivatives in the ODE/DAE system (in Eq. \eqref{eq:dae_final}) and convert the nonlinear system of ODEs/DAEs to a system of nonlinear algebraic equations. The solution to the resulting optimization problem is computed using an interior point solver.


\section{Computational Study} \label{sec:results}
We present results of applying our algorithm on a real data test case. The formulation in Section \ref{sec:state_param} is converted into a nonlinear program (NLP) using a finite difference approximation on the derivative terms. This technique of converting a continuous time problem to a finite-dimensional constrained NLP has been widely used in the optimal control literature \cite{Ross2003}, and has been applied previously in the context of gas pipelines \cite{Zlotnik2015}. We use a primal-dual interior point solver, IPOPT \cite{Wachter2009}, together with automatic differentiation in Julia/JuMP \cite{Dunning2017}, to compute the Jacobians, and solve the resulting NLPs. IPOPT is used because it leverages sparse linear algebra computations.  An error tolerance of $10^{-4}$ is used for all the computational experiments, which are done on a $2.9$ GHz, Intel Core i5 machine with $16$ GB RAM.


The test instance used to evaluate performance of the proposed computational estimation method is created using a capacity planning network model for an actual pipeline system and a month of time-series measurements from its supervisory control and data acquisition (SCADA) system.

\begin{figure}[ht!]
\centering
    \includegraphics[width=\linewidth]{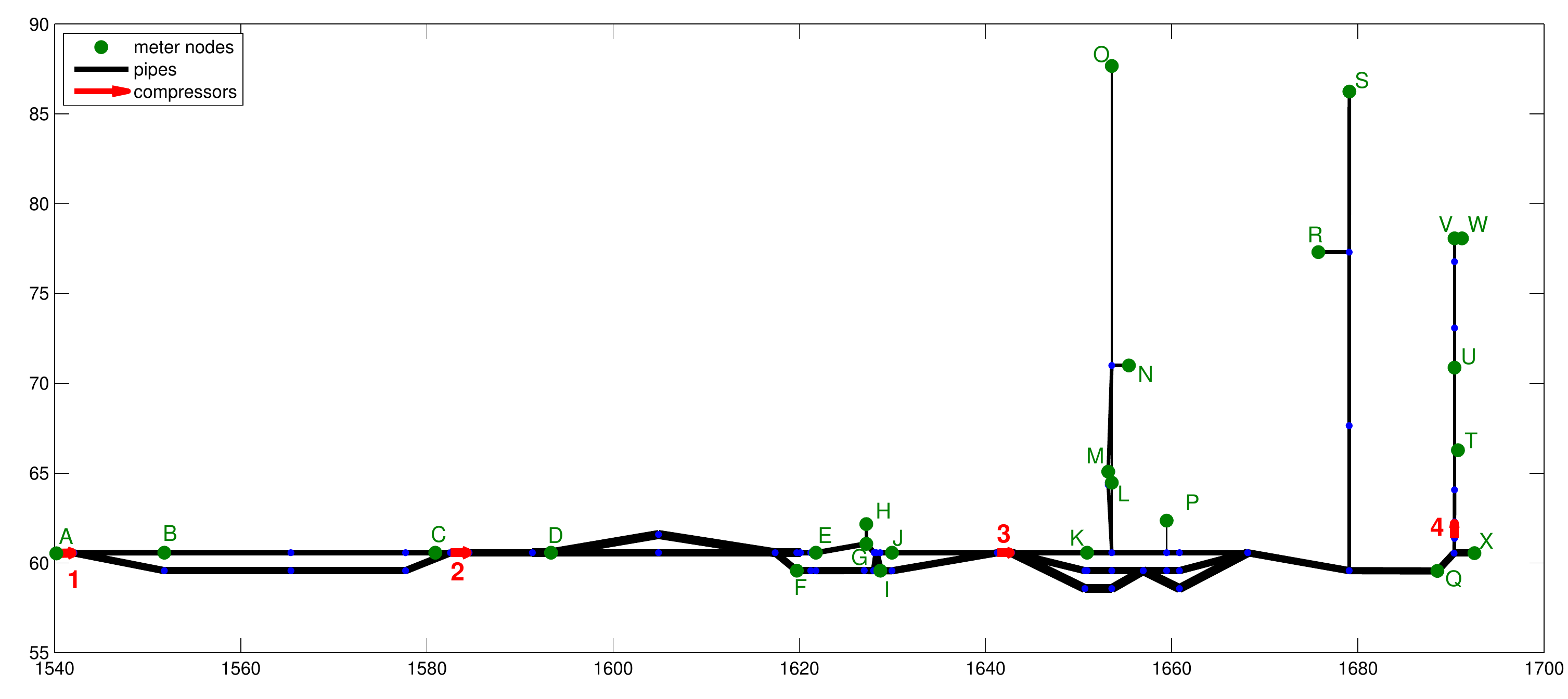}
    \caption{Schematic of pipeline subsystem. The green circles (labelled A to X) denote custodial meter stations.  Nodes without custodial meters are unlabelled. Red arrows (labelled 1 to 4) denote compressor stations.}
\label{fig:78node}
\end{figure}

\begin{figure}[ht!]
\centering
    \includegraphics[width=\linewidth, height=5.5cm]{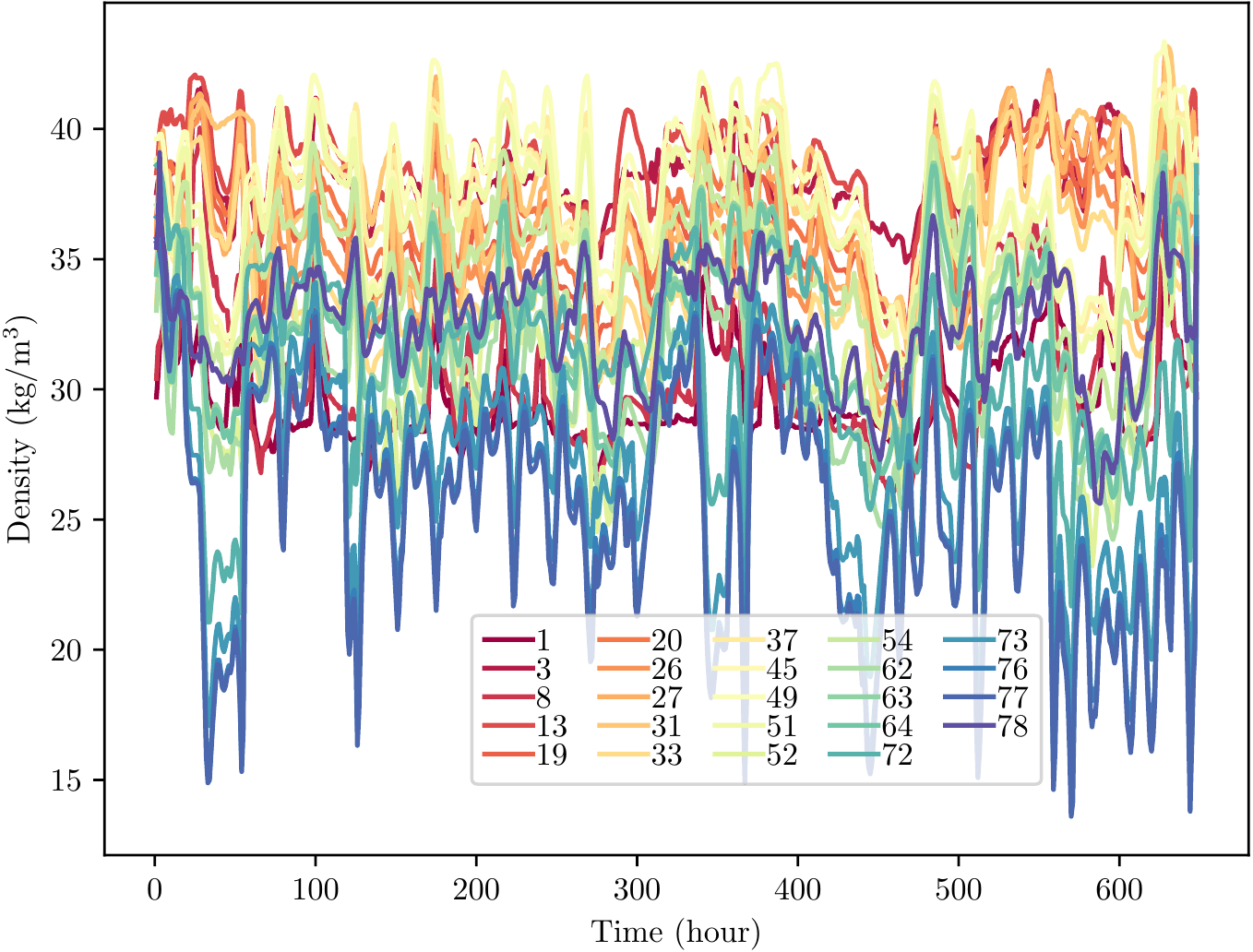}
    \caption{Time-series of density measurements at meter stations in Fig. \ref{fig:78node}.}
\label{fig:densities}
\end{figure}

The pipeline subsystem used is shown in Figure \ref{fig:78node}, and consists of 78 reduced model nodes, 95 pipes with total length of 444.25 miles, and 4 compressors. For each pipe, physical parameters are length, diameter, and friction factor given in the planning model. We use these friction factor values as ``ground truth'' to test the quality of our estimation algorithm. These values were scaled by an engineering factor of 1.1765 to compensate for pipe efficiency factors commonly used by commercial software packages but not considered in the reduced model approach.  This scaling value was determined in a simulation comparison study \cite{zlotnik2017psig}. The time-series data is taken from a SCADA system used for operation of the pipeline. This system provides hourly measurements of pressure, temperature, and volumetric flow out of the system at 31 metered custody transfer meter and check measurement locations, as well as average gas gravity and thermal content. Check measurements at the 4 compressor stations include pressure and temperature at suction and discharge, and volumetric through-flow. From this data, we used the CNGA equation of state \cite{menon05} to compute gas densities (shown in Figure \ref{fig:densities}) and withdrawals at network nodes, as well as compression ratios in the density variables, as described previously \cite{zlotnik2017psig}.
\vspace{-1ex}

\vspace{-3ex}

\vspace{-1ex}

\begin{figure}[t!]
    \centering
    \vspace{1ex}
    \includegraphics[width=\linewidth, height=4.5cm]{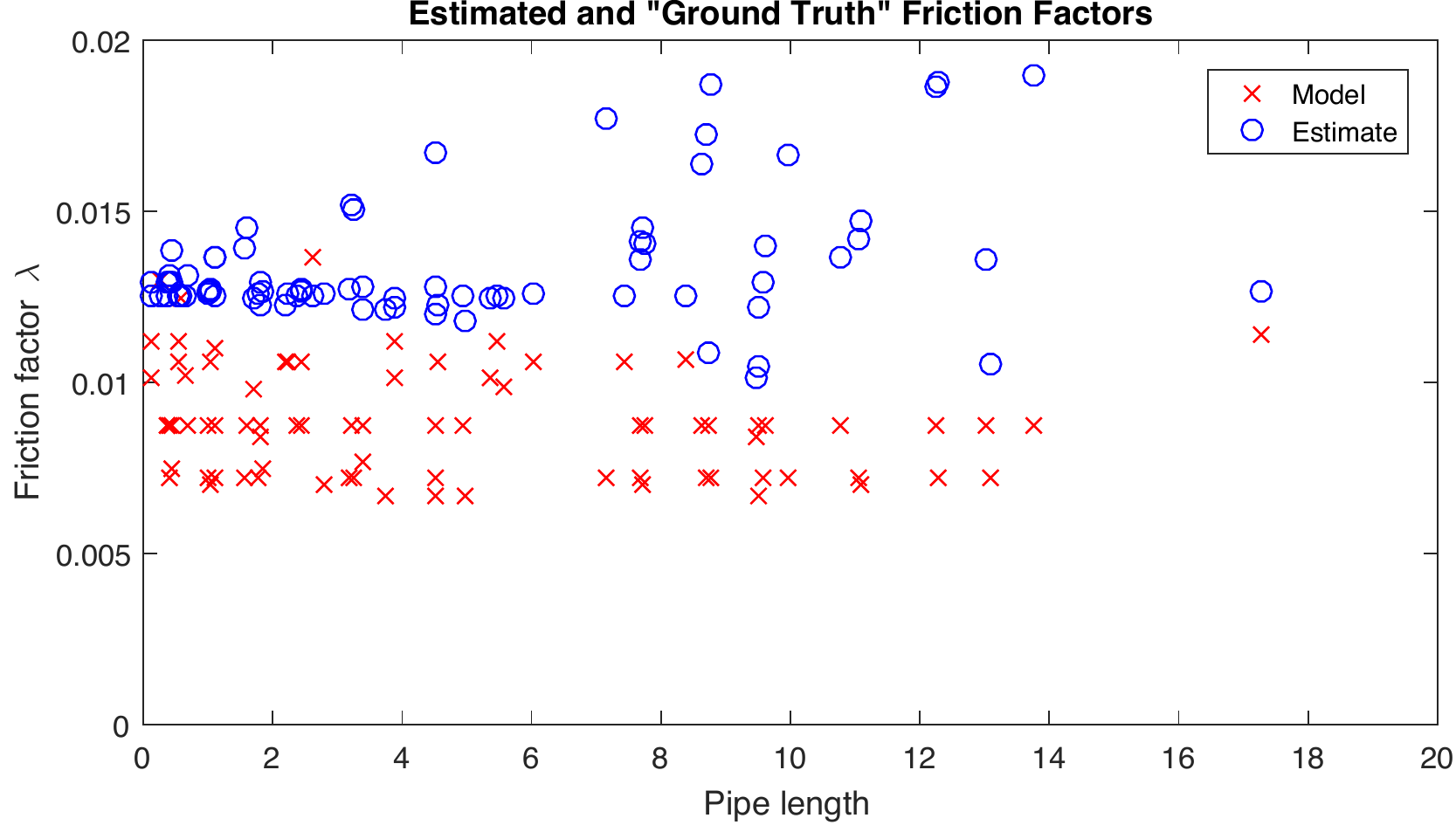}
    \caption{``Ground truth'' values (x) and estimates (o) of friction factors as a function of pipe length (in km).}
    \label{fig:estim}
\end{figure}

We solve the problem in \eqref{eq:state_2_sp} to yield estimates of the pressures and flows throughout the network, as well as of the friction factor parameters.  Because the flows and pressures throughout the system are unavailable in the data set, the only basis for evaluating the solution quality is to compare estimates of the friction factor values with those provided in the capacity planning model.  These values are plotted in Fig. \ref{fig:estim} for each pipe as a function of the pipe length.

In our previous study \cite{sundar2018}, the algorithms were tested on synthetic data on small test networks and with simulated pressure and withdrawal/injection measurements at all network nodes.  Here we are exposed to the challenges that arise in real data sets, where only time-series measurements are available at only a subset of nodes, the ``ground truth'' is approximate for model parameters and is not available for the state, and noise is present due to a variety of sources that are not characterized or may be unknown. As a result, we see that the friction factor estimates vary substantially from the values in the real data model.  Nevertheless, the identification procedure may result in a representation that could adequately be used for optimization and control.

\section{Conclusion} \label{sec:conclusion}
In this study we present formulations for joint state and parameter estimation for the flow of natural gas through a large-scale network of pipelines with actuation by compressors. We present approaches to physical and engineering modeling, model reduction, control system modeling, and uncertainty modeling.  The method has been used on synthetic data with pressure and flow measurements at all network nodes in our previous study \cite{sundar2018}.  Here we apply the method to a test case created from a capacity planning network model for an actual pipeline network and a month of time-series measurements from its supervisory control and data acquisition (SCADA) system.  Because the real data set has measurements at only 31 of the 78 network nodes, the parameter estimates show discrepancy from the values given in the planning model.

Model identification for large gas networks using only sparse pressure measurements thus remains an open challenge.  The modeling formalism and algorithmic approach presented here could be further developed and used in, for example, leak detection techniques for sparsely instrumented systems.  Future work will explore system identification for gas networks, and characterize bounds on the quality of state and parameter estimates. Estimation algorithms that are robust to outliers in the measurements will also be explored.  In addition, development of a Bayesian filtering approach for the DAE systems would be of general interest, and could be applied to gas network dynamic models for pipeline real-time state estimation and leak detection applications.


\section*{Acknowledgements}
This work was carried out as part of Project GECO for the Advanced Research Project Agency-Energy of the U.S. Department of Energy under Award No. DE-AR0000673. Work at Los Alamos National Laboratory was conducted under the auspices of the National Nuclear Security Administration of the U.S. Department of Energy under Contracts No. DEAC52-06NA25396 and 89233218CNA000001, and was supported by the Advanced Grid Modeling Research Program in the U.S. Department of Energy Office of Electricity.  The authors are grateful for support from Kinder Morgan Corporation, which provided the data used in this study.

\bibliographystyle{unsrt}
\bibliography{spe}

\end{document}